\documentclass[aps,pra, onecolumn,14 pts,a4paper]{revtex4}
 \pdfoutput=1

\usepackage{amssymb,amsmath,amsfonts,graphicx}
\usepackage{bm}
\usepackage{epsfig,color,times}
 \usepackage{bm,color}
\usepackage{graphicx}
\usepackage{amssymb}
\usepackage{epstopdf}

\begin{document}

\title{Blowing-up solutions of the axisymmetric Euler equations for an incompressible fluid}

\author{Yves Pomeau $^1$ and Martine Le Berre   $^2$}
\affiliation{ $^1$ Ladhyx, Ecole polytechnique, Palaiseau, France
\\$^2$  ISMO, Universit\'e Paris-Sud,  91405 Orsay Cedex, France}.

\date{\today }

\begin{abstract}
A 1934 paper by Leray posed the question of the regularity of solutions of the dynamical equations for incompressible inviscid fluids with smooth initial data. Since there has been many attempts to answer this question. Leray examined the possibility of self-similar solutions becoming singular in finite time at a definite space-time location. We reexamine this question in the light of a thorough analysis of the equations for the self-similar solution in axisymmetric geometries with a dependence on a logarithm of time besides the one due to the  transformation to self-similar variables.   
\end{abstract}

\maketitle

   \section{Introduction}
    \label{Introduction} 
    
    Our physical world is made partly of fluids like air and water. Beginning with Archimedes of Syracuse scientists have long tried to understand the dynamlcs of those fluids thanks to more or less elaborate mathematics. Euler in 1757  \cite{Euler} wrote one of the first ever partial differential equations, named now after him, for the motion of fluids without viscosity. This was completed later by Navier and Stokes to include the effect of viscosity. The effects of viscosity and of compressibility are measured by two numbers, the Mach number (for the incompressibility) and the Reynolds number (for the viscosity). Many natural and man-made flows have typically large to very large Reynolds number and negligible Mach number, the limit we shall deal with. The neglect of viscosity and compressibility leads  to the Euler equations, 
    \begin{equation}
{\partial_{t}{\bf{u}}} + {\bf{u}}\cdot \nabla {\bf{u}} = - \nabla p
\textrm{,}
\label{eq:Euler1}
\end{equation}
and 
\begin{equation}
\nabla \cdot {\bf{u}} = 0
\textrm{,}
\label{eq:Euler2.0}
\end{equation}
$\partial_{t}$ is for the time derivative, the vector  ${\bf{u}}({\bf{r}}, t)$ is the local value of the fluid velocity at time $t$, boldface being for vectors, and $p$ is the pressure, a gauge function allowing to satisfy the condition of incompressibility (\ref{eq:Euler2.0}). The nabla sign is for the gradient with respect to coordinate ${\bf{r}}$ and the mass density has been set to $1$.  

The next question is: what is predicted by Euler equations? 
 We do not know yet if smooth and bounded velocity fields in three space dimensions always yield a smooth solution at any later time. This is the question we address below and answer by the negative, at least within  some assumptions on the positiveness of  square modulus of solutions.
 We outline a method for getting by perturbation a solution diverging in finite time with smooth and bounded initial data.  This is based on the existence of non trivial explicit solutions, found by Bragg and Hawthorne (BH) \cite{BH}, of the steady axisymmetric Euler equation  (\ref{eq:Euler6ss})  with a parameter permitting to make the Leray advection term as small as wanted. This advection term  results from a mapping of the rest frame $({\bf{r}} ,t)$ for Euler equations into a frame $({\bf{R}},t)$ collapsing to a point in finite time, that gives the set of equations (\ref{eq:Euler1ss})-(\ref{eq:Euler2ss}),  called Euler-Leray equations below. Our goal is to find a solution of the Euler-Leray equations by perturbation describing the slow collapse of a solution of the steady Euler equation. 
 Our perturbative method introduces two solvability conditions in an expansion where the advection term is small. To be satisfied two solvability conditions require two free parameters, because the advection term has no free parameter and there is a priori no way to impose the vanishing of the integral in the solvability conditions at the first non trivial order of the expansion. Moreover there are reasons for excluding non trivial steady solutions of the Euler-Leray equations in an axisymmetric geometry. Free parameters are introduced into the game by considering a small perturbation to the steady Bragg-Hawthorne solution which oscillates in time. The expansion of the solution in powers of this small amplitude yields at second order a contribution to the solvability conditions which, once added to the one due to the small advection term, yields the requested two free parameters if there are two oscillation modes. This is the general principle of our solution by perturbation.

In 1934 Leray \cite{leray} suggested to explain the irregularity of turbulent flows by the loss of predictability of the evolution problem linked to the occurrence of singularities. Such singularities would forbid to continue the solution afterwards within the framework of deterministic fluid equations. Leray went further than that and wrote down the equation for self-similar solutions of the Navier-Stokes equation (including the viscosity - see below) describing the evolution from smooth initial data to a solution singular at one point of space and time (his equation (3.11)).  This paper is devoted to the search of solution of Leray's equations without viscosity which amounts to find singular solution of Euler equations. 

A major question in the search of self-similar solution of the Euler equations is how to take into account the invariants of those equations. There are two sets of invariants (excluding linear and angular momentum and helicity). First we have conservation of energy $ {\mathcal{E}} =  \frac{1}{2} \int \mathrm{d}{{\bf{r}}}  \ {\bf{u}}^2 \textrm{.}$ 
If this integral converges, it defines a positive quantity which is constant in the course of time when ${\bf{u}}({\bf{r}}, t)$ is a smooth solution of the Euler equations. There are also infinitely many conserved quantities which are the integrals $\int \mathrm{d} {\bf{s}} \cdot {\bf{u}} $ along any closed line of element $\mathrm{d} {\bf{s}}$, this line being carried by the flow. This Kelvin circulation theorem, a  highly non -trivial property, has important consequences in the search for self-similar solution of the Euler equation: if a closed line is carried by the flow inside the collapsing domain, Kelvin theorem constrains the possible exponents of the similarity solution and, as we shall see, this constraint is not compatible (for singularities occurring at a single point) with the conservation of energy inside the collapsing domain. 
The exponents derived from the conservation of circulation and the ones derived from the conservation of energy are different. Therefore one has, somehow, to choose which one is right. As argued in reference  \cite{CR} it is not possible to use the conservation of energy to build up a self-similar solution if energy is given by a diverging integral. In this case the only possibility is to use the conservation of circulation. This divergence is present when looking for a steady solution of Euler-Leray in the general 3D case because the velocity field decays too slowly at large distance to make the energy finite.

Therefore, if one chooses instead the conservation of energy to derive the scaling laws of the self-similar solution, one finds that the energy of a steady solution is formally diverging logarithmically at large distances, when considered for a steady solution. However we shall not deal with a steady solution, but with a solution that is localized and decays quickly at large distances with a small added time dependent perturbation. 

In the geometry we shall consider two other invariants could constrain the similarity solution, the linear momentum and the angular momentum along the symmetry axis. The stream function we shall find is odd with respect to $z$, which makes vanish both momenta, which are therefore irrelevant as parameters of the scaling of the similarity solution. 

 As did Leray \cite{leray} let us suppose there is a singularity in finite time by the evolution of the flow and that this singularity is of the self-similar type. The corresponding solution of the Euler equations reads  $$ {\bf{u}}( {\bf{r}}, t) = (t^*- t)^{-\alpha}  {\bf{U}} ( {\bf{r}}(t^*- t)^{-\beta})\textrm{,}$$ where $t^*$ is the time of the singularity (set to zero afterwards), where $\alpha$ and $\beta$ are positive exponents to be found and where  $ {\bf{U}}(.)$ is to be derived by solving Euler equations. 
 
 That such a velocity field is a solution of Euler equations implies $ 1 = \alpha + \beta$. The conservation of circulation implies $ 0 = \alpha - \beta$, and therefore $\alpha = \beta = 1/2$. 
 
 The conservation of the two Noether invariants of angular and linear momentum could lead to other constraints on the exponents. Likewise the energy, they are formally not constant if one takes the scaling exponents $\alpha = \beta = 1/2$. But, contrary to the energy, those invariants can be set to zero for a non zero velocity field. Another possibility is to take them equal to infinity in the similarity solution. This agrees with the slow decay  of the velocity field at large distances from the singularity point, like $1/r$, which makes diverge both momenta. We shall assume that, like the energy, this divergence implies that the scaling exponents of the self-similar solution are not determined by the conservation of those momenta. 
 With the stretched variable ${\bf{R}} =  {\bf{r}}(- t)^{-\beta}$,  the Euler equations become a set of equations (the Euler-Leray equations) for $ {\bf{U}}({\bf{R}})$: 
 \begin{equation}
- ( \alpha  \ {\bf{U}}  + \beta  \ {\bf{R}}\cdot \nabla {\bf{U}}) + {\bf{U}}\cdot \nabla {\bf{U}} + \nabla P = 0
\textrm{,}
\label{eq:Euler1ss}
\end{equation}
\begin{equation}
\nabla \cdot {\bf{U}} = 0
\textrm{.}
\label{eq:Euler2ss}
\end{equation}
For a reason explained later we shall work with an extended version of the similarity equation including an explicit time dependent part. This is done by using as variable $\tau = - \ln(t^* - t)$ and by keeping the velocity field depending on  ${\bf{r}}$ via the stretched radius   ${\bf{R}} $. Together with equation (\ref{eq:Euler2ss}) this makes the modified Euler-Leray equation:  
 \begin{equation}
\frac{\partial {\bf{U}}}{\partial \tau} - ( \alpha  \ {\bf{U}}  + \beta  \ {\bf{R}}\cdot \nabla {\bf{U}}) + {\bf{U}}\cdot \nabla {\bf{U}} + \nabla P = 0
\textrm{,}
\label{eq:Euler3sm}
\end{equation}
We shall consider below an axisymmetric geometry with swirl and try to find a tractable solution for the above Euler-Leray equations in  their modified form (\ref{eq:Euler3sm})-(\ref{eq:Euler2ss}) including the time variable $\tau$.  In a first step, presented in  \ref{BHeq}-\ref{localized}, we drop the linear part of (\ref{eq:Euler3sm}), reducing it to a Bragg-Hawthorne equation for the stream fonction. This assumes that the linear part is much smaller than the nonlinear part. In a second step, treating the linear part as perturbation, we explain in  \ref{lackperturbed}-\ref{perturbed} why no ''smooth solution''  of  the steady Euler-Leray equations (\ref{eq:Euler1ss})-(\ref{eq:Euler2ss}) exists, a result already stated in \cite{YP}.  Differently  a solvability condition at first order can be found for the time-dependent  equations (\ref{eq:Euler3sm})-(\ref{eq:Euler2ss}), allowing to build a solution oscillating periodically with respect to time $\tau$, as  shown in \ref{perturbed.1}.

That the time oscillations are undamped if they are of small amplitude is not trivial. This lacks of damping follows in general from the property that the dynamical equations conserve the energy. In the case of equation (\ref{eq:Euler3sm}) this property is a consequence of the choice of the Sedov-Taylor exponents as it can be expected. The proof goes as follows. The terms $ ({\bf{U}}\cdot \nabla {\bf{U}} + \nabla P )$ conserve the energy $  \frac{1}{2} \int { \mathrm{d}} {\bf{R}}  {\bf{U}}_i  {\bf{U}}_i $ where the convention of summation on the same coordinate indices is applied. There remains to prove the following property for $( \alpha  \ {\bf{U}}  + \beta  \ {\bf{R}}\cdot \nabla {\bf{U}}) $ when the Sedov-Taylor exponents are used. For that purpose one has to show that, for any smooth incompressible velocity field,  
 \begin{equation}
  \int { \mathrm{d}} {\bf{R}}  \ {\bf{U}}_i ( \alpha  \ {\bf{U}}_i +  \beta  \ {\bf{R}}_j\partial_j  {\bf{U}}_i ) = 0 
  \textrm{.}
\label{eq:but}
\end{equation}
Integrating by part one obtains  
$$  \int { \mathrm{d}} {\bf{R}}  \ {\bf{U}}_i \ {\bf{R}}_j\partial_j  {\bf{U}}_i  = - D  \int { \mathrm{d}} {\bf{R}}  \ {\bf{U}}_i {\bf{U}}_i -  \int { \mathrm{d}} {\bf{R}}  \ {\bf{U}}_i \ {\bf{R}}_j\partial_j  {\bf{U}}_i   \textrm{,}$$
where $D$ is the physical dimension of space ($D = 3$ in the present case). Therefore one has
 \begin{equation}
  \int { \mathrm{d}} {\bf{R}}  \ {\bf{U}}_i \ {\bf{R}}_j\partial_j  {\bf{U}}_i  = -  \frac{D}{2}  \int { \mathrm{d}} {\bf{R}}  \ {\bf{U}}_i {\bf{U}}_i   \textrm{.}
\label{eq:but2}
\end{equation}   
 and 
 $$  \int { \mathrm{d}} {\bf{R}}  \ {\bf{U}}_i ( \alpha  \ {\bf{U}}_i +  \beta  \ {\bf{R}}_j\partial_j  {\bf{U}}_i )  =  \int { \mathrm{d}} {\bf{R}} \ {\bf{U}}_i {\bf{U}}_i  (\alpha  - \frac{D \beta}{2}) \textrm{.}$$
 With the other relation $ \alpha + \beta = 1$,  this yields 
 \begin{equation}
  \alpha = \frac{D}{2 + D}  \qquad \text{and}  \qquad \beta = \frac{2}{2 + D}  \textrm{,}
 \label{eq:Sedovexp}
\end{equation}   
 which are the Sedov-Taylor exponents.

  \section{Euler-Leray equations in the axisymmetric case}
    \label{solaxis} 

The solution of equations (\ref{eq:Euler3sm}) makes a highly non trivial problem. Therefore it is desirable to transform them to simplify their formulation and hopefully to map them into a tractable numerical/analytical problem. This is done in two steps. First, without the first two terms, 
 \begin{equation}
\frac{\partial {\bf{U}}}{\partial \tau} - ( \alpha  \ {\bf{U}}  + \beta  \ {\bf{R}}\cdot \nabla {\bf{U}}) 
 \textrm{,}
 \label{eq:linear}
\end{equation}   
equation (\ref{eq:Euler3sm}) becomes the steady Euler equations,
 \begin{equation}
 {\bf{U}}\cdot \nabla {\bf{U}} + \nabla P = 0 \textrm{.}
 \label{eq:steadyEuler}
\end{equation}   
 A number of tricks allow to make this last equation far simpler in the case of axisymmetric flows with a swirl, namely with no dependence of the velocity and pressure with respect to the azimuthal angle $\phi$ but with a non-vanishing azimuthal component of the velocity. This allows to map the equation into a single elliptic equation for the stream function in a half plane bounded by the axis of symmetry, an equation derived first by Bragg and Hawthorne \cite{BH}. This equation is derived below. Of course this does not yield directly a solution of the Euler-Leray equations because it does not take into account what we shall call the advection term, namely $\ - ( \alpha  \ {\bf{U}}  + \beta  \ {\bf{R}}\cdot \nabla {\bf{U}}))$ as it appears on the left hand side of equation (\ref{eq:Euler1ss}). Nevertheless one notices that the Bragg-Hawthrorne solution concerns the  {\it{nonlinear}} part of the Euler-Leray equations only. Therefore it is possible to make the linear part small with respect to the rest of the equation by taking a BH  solution of "large modulus" in a sense to be made more precise and by expanding the solution of Euler-Leray equations in the limit of a linear advection term at first order with respect to the small parameter. 
However we show below that this cannot yield a non trivial solution of the Euler-Leray equations without dependence with respect to the time $ \tau$. 

As the equation for  $U_\phi$ is the same as the ones for  $u_\phi$, but with a positive divergence added, it predicts that $U_\phi$ is carried along diverging spirals in the plane $(R, Z)$ making it impossible to have a steady solution decaying to zero at infinity. However this does not make it impossible to have a solution of the time dependent equation (\ref{eq:Euler3sm}). Such a non trivial solution can be build as follows. We still consider a solution of the axisymmetric Euler equation with a velocity of very large magnitude so that the terms in (\ref{eq:linear}), which are the derivative with respect to time $\tau$ and the streaming term,  can be seen as small (but not irrelevant!) perturbations. Time dependence  introduced at first order leads to small amplitude oscillations. Somehow this allows to balance the stretching due to the advection term by a kind of small Stokes mean drift due to the oscillations. 

The introduction of time oscillations in our problem can be seen as an application to the present situation of one of the ideas expressed by Kolmogorov in his 1941 paper on turbulence \cite{kolm}. 
A careful reading of this paper shows that Kolmogorov had a qualitative idea of the cascade that was not without relationship with the present idea of oscillating self-similar solution. In the translated version one finds an image of the cascade with a clear time dependence in the energy transfer from large scale to smaller scale. Kolmogorov sees this transfer as occurring by instabilities developing at a smaller scale that has been created by the decay of fluctuations at a bigger scales. Even though Kolmogorov never refers to Leray, there is nothing against the application of this idea of oscillations leading to smaller and smaller scales to the self-similar solution of the problem imagined by Leray.  Somehow the oscillations could be seen as a way to cope with the outside drift due to the contraction of scale induced by Leray's change of coordinates . 

The idea of oscillations besides the regular self similar dynamics has been present in the theory of turbulence in various forms. We refer the interested reader to the pioneering paper by Pumir and Siggia \cite{SP}.

 \subsection{Bragg-Hawthorne equation}
    \label{BHeq}
    Let us write the Euler equations in the  axisymmetric case, namely for a velocity field written in cylindrical coordinates $(r, z, \phi)$ with $r$ distance to the $z$ axis, and $\phi$ azimuthal angle in the plane perpendicular to this $z$ axis. The components of the velocity field  are $u_r$, $u_z$ and $u_\phi$, all depending on $(r, z)$ only. 
    
    The incompressibility condition reads: 
\begin{equation}
  \frac{\partial (r u_r)}{\partial r} +  \frac{\partial (r u_z)}{\partial z} = 0
\textrm{.}
\label{eq:Eulerinc}
\end{equation}
It is satisfied by introducing the stream function $\Psi(r, z)$ such that 
\begin{equation}
u_r =   \frac{1}{r}   \frac{\partial  \Psi}{\partial z}
\textrm{.}
\label{eq:ur-psi}
\end{equation}
and 
 \begin{equation}
u_z =  -  \frac{1}{r}   \frac{\partial  \Psi}{\partial r}
\textrm{.}
\label{eq:uz-psi}
\end{equation}
  For a steady state  the Euler equation for $u_\phi (r, z)$ is,
 \begin{equation}
u_r \frac{\partial u_\phi}{\partial r}  + u_z \frac{\partial u_\phi}{\partial z}   +  \frac{u_r u_\phi}{r}= 0
\textrm{,}
\label{eq:Euler4ss}
\end{equation}
This equation is remarkable because it is linear homogeneous with respect to $u_\phi(r, z)$. Moreover it can be integrated, as a consequence of the conservation of circulation on closed curves circling the symmetry axis. It can be rewritten as
 \begin{equation}
 u_r \frac{\partial (r u_\phi)}{\partial r}  + u_z \frac{\partial (r u_\phi)}{\partial z} = 0
\textrm{,}
\label{eq:Euler3ss.1}
\end{equation}
This shows that $(r u_\phi)$ is a function of $\Psi$, denoted usually like $B(\Psi)$: 
 \begin{equation}
  u_\phi= \frac{B(\Psi)}{r}
\textrm{,}
\label{eq:uphi-psi}
\end{equation}

The BH equation is derived by coming back to the general form of the steady Euler equations: 
 \begin{equation}
 {\bf{u}}\cdot\nabla{\bf{u}} + \nabla p = 0
\textrm{,}
\label{eq:Euler6ss}
\end{equation}
This is equivalent to: 
 \begin{equation}
 \nabla H -   {\bf{u}}\times {\bf{\omega}} = 0 \textrm{,}
\label{eq:Euler7ss}
\end{equation}

where ${\bf{\omega}} = \nabla \times  {\bf{u}}$ is the vorticity and where $H(\Psi) = \frac{1}{2} u^2 + p$ depends on $\Psi$ only as it follows by projecting this equation on the flow lines. 

Consider now the component of equation (\ref{eq:Euler7ss}) in the $z$ direction: 
 \begin{equation}
\frac{{\mathrm{d}}H}{{\mathrm{d}}\Psi}\frac{\partial \Psi}{\partial z} - u_r \omega_{ \phi} +  u_\phi \omega_{r } = 0
\label{eq:Euler8ss}
\end{equation}
Consider the last two terms in this equation. From its definition,  $\omega_{ \phi} = \frac{1}{r} \nabla^2 (r \Psi) = \frac{1}{r}  (\frac{\partial^2 (r \Psi)}{\partial z^2} + \frac{\partial^2 (r \Psi)}{\partial r^2})$. In the same way one finds $\omega_{r }  = - \frac{\partial u_{\phi}}{\partial z} =- \frac{{\mathrm{d}}B}{{\mathrm{d}}\Psi}\frac{\partial \Psi}{\partial z}$ and $u_r = - \frac{1}{r} \frac{\partial \Psi}{\partial z}$.  Therefore all terms in equation (\ref{eq:Euler8ss}) bear a factor $\frac{\partial \Psi}{\partial z}$. Once this is factored out one obtains the Bragg-Hawthorne equation:
\begin{equation}
{\mathcal{L}} \Psi - r^2 \frac{{\mathrm{d}}H}{{\mathrm{d}}\Psi} +  B  \frac{{\mathrm{d}} B}{{\mathrm{d}}\Psi} = 0
\textrm{,}
\label{eq:Euler9ss}
\end{equation}
with ${\mathcal{L}}  = \frac{\partial ^2}{\partial z^2 } + \frac{\partial ^2}{\partial r^2 } - \frac{1}{r} \frac{\partial}{\partial r} \textrm{.}$

Various solutions of this equation have been found, either numerical or -partly- analytical. We look for a solution that is steady with a fluid at rest at infinity and without boundary, not the case of most solutions  of the literature. This makes the perturbation calculation done later more straightforward without the need to take into account an average velocity. Such a steady solution can follow from an imposed reflection symmetry: the velocity field is assumed to be mirror symmetric with respect to the plane $ z = 0$. It means that component $u_z$  of the velocity is an odd functions of $z$ and that $u_r$ is an even function of $z$. Because $u_{\phi}$ is proportional to $\Psi$ with our choice of $H(\Psi)$,  $u_{\phi}$ is also an odd function of $z$, which cancels to zero the angular momentum of the solution along the vertical axis. That $u_z$ is odd with respect to $z$  implies straight away that a solution having this symmetry have no mean velocity in the $z$ direction: if it had one, this should be reversed by the mirror symmetry while remaining identical. Because the solution itself is mirror symmetric this average velocity along $z$ can be only zero. This yields a Dirichlet boundary condition on the plane $z = 0$ for the solution of the BH equation. With this choice, the function $\Psi$ is an odd function of $z$. This imposes constraints on the functions $H$ and $B$. Because ${\mathcal{L}} \Psi$ is odd, any function in equation (\ref{eq:Euler9ss}) must be odd as well. This implies that $H(\Psi)$ is an even function of $\Psi$ while $B(\Psi)$ is odd.  A simple way to ensure that is to have $H(\Psi)$ quadratic and $B(\Psi)$ linear: 
\begin{equation}
 H(\Psi) = \frac{\gamma}{2} \Psi^2 \textrm{,}
\label{eq:H}
\end{equation}

and 
\begin{equation}
 B (\Psi) = \delta \Psi 
 \textrm{,}
\label{eq:B}
\end{equation}

with $\gamma$ and $ \delta$ free constants at this stage. 
With this choice the BH equation  becomes the  {\emph{linear}} homogeneous equation,
\begin{equation}
 {\mathcal{L}} \Psi -  \gamma r^2 \Psi  +   \delta^{2} \Psi = 0 
  \textrm{,}
\label{eq:Lpsi}
\end{equation}
Because the coefficients of this equation are independent on the variable $z$ one can assume that the solution depends on $z$ as $e^{ik z}$ with $k$ real. Solving for an arbitrary $k$ one obtains the general solution by adding solutions for different values of $k$. Let  $\Psi =  e^{ik z} \varphi_k(r)$ be a solution associated to the wavenumber $k$. The function $\varphi_k(r)$ is the solution of the linear ordinary differential equation, 
\begin{equation}
\frac{{\mathrm{d}^2}\varphi_k(r)}{{\mathrm{d}}r^2 } - \frac{1}{r} \frac{{\mathrm{d}}\varphi_k(r)}{{\mathrm{d}}r}  -  \gamma r^2  \varphi_k(r) +  ( \delta^{2} - k^2)  \varphi_k(r) = 0
\textrm{,}
\label{eq:Euler10ss}
\end{equation}
This equation is posed on the half-real line $r\geq 0$. Therefore boundary conditions are to be imposed for $r = 0$ and $r$ tending to infinity.  Consider first the case $r = 0$.  To ensure the smoothness of the velocity field on the axis $r = 0$, the function $ \varphi_k(r)$ must have a Laurent expansion starting like 
\begin{equation}
  \varphi_k(r) =  c ( 1 +  \frac{k^{2}- \delta^{2}} {2}  r^2   \ln(r) + ...) + d  \ \ r^{2} \left( 1 +  \frac{k^{2}- \delta^{2}} {8}  r^2 +... \right)
  \textrm{,}
\label{eq:c}
\end{equation}
 where $c$ and $d$ are arbitrary constants. From the equation the coefficients  of the Laurent  expansion  for the two independent solutions can be computed recursively.

 \subsection{Asymptotic behavior of solutions of (\ref{eq:Euler10ss}) and (\ref{eq:Lpsi}), search of a localized solution}
 \label{localized}
The behavior of solutions of (\ref{eq:Lpsi}) at $r$ and $z$ tending to infinity is less easy to find than the behavior near $r = 0$. At leading order the solution of equation (\ref{eq:Euler10ss}) is given by a function with an exponential factor (what is called sometimes the WKB approximation) like 
\begin{equation}
  \varphi_k(r) \approx A(r) e^{S(r)} \textrm{.}
\label{eq:phik}
\end{equation}  
Putting this form of  $\varphi_k(r)$ into equation (\ref{eq:Euler10ss}) and keeping the leading order term only (at large $r$) one finds 
\begin{equation}
 S'^2(r) = \gamma r^2 \textrm{,}
 \label{eq:Sprime2}
\end{equation}  
 where $S'$ is the first derivative of $S(r)$ with respect to $r$. 
If $\gamma$ is positive, $S'(r)$ is $\pm  \gamma^{1/2} r$. Therefore $e^{S(r)}$ either grows or decay exponentially as $r$ tends to infinity. Because the solution given by the Laurent expansion (\ref{eq:c}) depends on two parameters and because equation  (\ref{eq:Euler10ss})  is linear and homogeneous, the amplitudes of the exponentially growing and decaying modes are related to $c$ and $d$ by a linear matrix.  We shall assume that  this matrix has a non zero determinant, that is justified in particular because this determinant depends continuously on the parameters $\delta$ and $\gamma$ which are not fixed. It follows that  $c$ and $d$, as defined by equation (\ref{eq:c}), can be chosen, up to a trivial multiplicative factor, such that the coefficient of the exponentially growing mode is cancelled. Therefore, with a convenient choice of the ratio $c/d$, there is a solution of (\ref{eq:Euler10ss}) decaying like $e^{- \frac{1}{2} \gamma^{1/2} r^2}$ as $r$ tends to infinity. This is the solution we shall deal with now.

Adding contributions with different values of $k$ one may obtain a whole set of solutions of (\ref{eq:Lpsi}) decaying as well in the vertical direction. 
A simple choice of weight function of possible values of $k$ is a Gaussian times an arbitrary  polynomial $P(k)$ which is odd with respect to $k$ to make the integral over $k$ non zero. This  yields for $\Psi (r, z)$:  

\begin{equation}
\Psi = \int_{-\infty}^{+\infty} {\mathrm{d}}k \ e^{-(k r_0)^2}  P(k) \sin(k z) \ \varphi_k(r) 
\textrm{,}
\label{eq:Euler11ss}
\end{equation}
The behavior of $\Psi$ as $z$ tends to infinity is derived by the saddle point method combining the exponential coming from the Gaussian weight $e^{-(k r_0)^2}$ and the one coming from the sinus function. It is given by 
$$ \Psi  \sim e^{- \frac{z^2}{4 r_0^2}} \textrm{.}$$
This solution of the BH equations is localized and it is obviously a square integrable function, something which matters because we are going to deal with integrals of product of components of the velocity field. 

The  solution in equation (\ref{eq:Euler11ss})  has been written without normalization factor. Such a factor is irrelevant in the present discussion because the amplitude of $\varphi_k(r)$, solution of a linear homogeneous equation, is arbitrary. By tuning the various free parameters in this solution, one may give arbitrary values of the order of magnitude of the stream function and of the length scale of this BH velocity field. Thanks to that one may manage to make the advection term in the Euler-Leray equation as small as wished with respect to the regular Euler part of this equation.  

It is of interest to understand the physical meaning of this limit (Leray's advection term small with respect to the non linear term in equation (\ref{eq:Euler1ss})). Looking at the scaling laws for the collapse one sees that the order of magnitude of the ratio of the nonlinear term to the  linear term is of order $(\tilde{\omega}t)$, where $\tilde{\omega}$ is the order of magnitude of the vorticity in the flow at the starting time, whereas $t$ is the time it takes to reach the singularity from this starting time. The condition that the advection term is small (but non zero) is $\tilde{\omega} t \gg 1$. In other terms the intrinsic dynamical time for the flow configuration, namely $1/\tilde{\omega}$ is much shorter than the time until blow up. This is consistent with the idea that this blow-up is linked to a small departure of the initial condition from a steady Bragg-Hawthorne solution: if the initial condition is exactly BH, it stays so forever and no blow-up takes place, whereas a slight departure yields a blow-up after some time, proportional to the inverse of the magnitude of this initial departure.

 \subsection{Lack of smooth time independent solution of axisymmetric Euler-Leray equations}
 \label{lackperturbed}
 Let us come back to equation (\ref{eq:Euler4ss}) for the azimuthal component of the velocity field and add to it the advection term by the Leray change of scale. This gives: 
  \begin{equation}
\frac{1}{2}(u_\phi + r \frac{\partial u_\phi}{\partial r} + z \frac{\partial u_\phi}{\partial z}) + u_r \frac{\partial u_\phi}{\partial r}  + u_z \frac{\partial u_\phi}{\partial z}   +  \frac{u_r u_\phi}{r}= 0
\textrm{,}
\label{eq:Euler5sm}
\end{equation}
The first term on the left-hand side, namely $\frac{1}{2}(u_\phi + r \frac{\partial u_\phi}{\partial r} + z \frac{\partial u_\phi}{\partial z}) $ can be put into the next two terms by introducing the "velocity" field of components,
 \begin{equation}
 w_z = u_z +  \frac{z}{2} \textrm{,}
 \label{eq:wz}
\end{equation}
and 
 \begin{equation}
  w_r = u_r +  \frac{r}{2} \textrm{.}
   \label{eq:wr}
\end{equation}
With the help of this new velocity field equation (\ref{eq:Euler5sm}) can be written as: 
 \begin{equation}
 w_r \frac{\partial( ru_\phi)}{\partial r}  + w_z \frac{\partial (r u_\phi)}{\partial z} = 0
\textrm{,}
\label{eq:Euler4sm}
\end{equation}
It shows that the quantity $ru_\phi$ is convected along the flow lines of the velocity field $(w_r, w_z)$. Because this flow has constant divergence, $3/2$ with our scalings, the flow lines spiral out to infinity, whereas  if the divergence was equal to zero, the flow lines would make a  set of closed lines nested around a fixed point. If the perturbation by the Leray term is small, the divergence will be felt at large distances only. Therefore the only fixed point where $ru_\phi$ can take a non zero value whilst tending to zero at infinity, is a Dirac delta function at the fixed point of the $w$ field. But such  delta function is excluded because we are looking for a solution of Euler-Leray's equation which is smooth and of bounded amplitude. 

This absence of solution of  {\emph{steady}} Euler-Leray's equations with swirl explains why we have to look at the equations with the log-time dependence written in equation (\ref{eq:Euler3sm}). More specifically we shall look at a solution with a oscillatory dependence upon the logarithmic time $\tau$. As already said such an  oscillatory dependence is at least implicit in Kolmogorov's picture of cascade.

 \section{Perturbation of a Bragg-Hawthorne velocity field leading to a singularity}
 \label{perturbed}
Assuming that $\lambda$ is large, we shall expand the solution of the modified Euler-Leray equations  (\ref{eq:Euler2ss})-(\ref{eq:Euler3sm}) with respect to the small parameter $1/\lambda$. At leading order $\lambda^{2}$ of (\ref{eq:Euler3sm}),  the modified Euler-Leray equations reduce to the stationary Euler equations (\ref{eq:steadyEuler}), then we have,
   \begin{equation}
{\bf{U}} \approx \lambda {\bf{U}}^{(0)}  \qquad P  \approx \lambda^{2} P^{(0)}
\textrm{,}
\label{eq:leading1}
\end{equation}
where  the set $\left( {\bf{U}}^{(0)}, P^{(0)}\right)$ is a finite and fixed amplitude solution which satisfies the equations,
    \begin{equation}
{\bf{U}}^{(0)} \cdot \nabla {\bf{U}}^{(0)}+ \nabla  P^{(0)} =0 \qquad  \nabla \cdot {\bf{U}}^{(0)} = 0
\textrm{.}
\label{eq:leading2}
\end{equation}
Putting this leading order solution in the modified Euler-Leray equations, we shall prove that one obtains an expansion of  this solution in inverse powers of  $\lambda$, that gives for the velocity,
  \begin{equation}
 {\bf{U}} = \lambda {\bf{U}}^{(0)} +  \lambda^{1/2} {\bf{U}}^{(1)} +  {\bf{U}}^{(2)} + .... \textrm{,}
\label{eq:expansion}
\end{equation}
where all the  ${\bf{U}}^{(i)}$'s are of order one. In this expansion, the successive terms can be computed recursively, with  the constraint of solvability conditions. 

Let us now expand (\ref{eq:Euler3sm})  at order $\lambda^{3/2}$. We assume that the log-time scales as $\tau=\tau^{(1)}/\lambda$, which will be justified later.  At  order  $\lambda^{3/2}$  of (\ref{eq:Euler3sm}) we  have  to include the time derivative term  and to linearize the nonlinear part  of the equation around the solution $ {\bf{U}}^{0} $. For convenience we introduce the  linear operator ${\mathcal{M}}_{0} $,
\begin{equation}
{\mathcal{M}}_{0}  [ {\bf{U}}^{(1) },P^{(1)} ]=  {\bf{U}}^{(1)}\cdot \nabla {\bf{U}}^{(0)} + {\bf{U}}^{(0)}\cdot \nabla {{\bf{U}}}^{(1)} +  \nabla P^{(1)} \textrm{,}   
\label{eq:Mo}
\end{equation}
where the pressure $ P^{(1)} $ allows to satisfy the incompressibility condition $ \nabla \cdot {\bf{U}}^{(1)} =0$.   When we put the  term $ \lambda^{1/2}{\bf{U}}^{(1)}$ in the modified Euler-Leray equation (\ref{eq:Euler3sm}), we get , at order $\lambda^{3/2}$, the equation
\begin{equation}
 \frac{\partial {\bf{U}}^{(1)}}{\partial \tau^{(1)}}+ {\mathcal{M}}_{0}  [ {\bf{U}}^{(1) },P^{(1)} ]=0
 \textrm{,}   
\label{eq:dtau1}
\end{equation}
which can be solved directly without adding a solvability condition.
This result is a bit similar to Rayleigh linear equation for the stability of parallel flows, but in the present case it is far more complicated because of the dependence in space of the BH flow field. Equation (\ref{eq:dtau1}) for the velocity field is autonomous with respect to time $\tau^{(1)}$, and  linear with coefficients independent of $\tau^{(1)}$. The solution  of (\ref{eq:dtau1}) is a linear superposition of functions of space multiplied by the exponential factor $e^{\sigma \tau^{(1)}}$ where $\sigma$
is to be found by solving the eigenvalue problem, 
\begin{equation}
{\mathcal{M}}_{0} [ {\bf{U}}^{(1) },P^{(1)} ]+\sigma {\bf{U}}^{(1)} = 0  \qquad \nabla \cdot {\bf{U}}^{(1)} =0
 \textrm{,}
\label{eq:U1}
\end{equation}
where $\sigma$ is of order unity.  
Note that $\sigma = 0$ is a double eigenvalue of the operator ${\mathcal{M}}_{0}$ as follows from the invariance of the Euler equation under dilation of its amplitude and of its argument.  Moreover  non zero eigenvalues $\sigma$, if they exist, are purely imaginary because of the conservation of energy. We assume that complex conjugate eigenvalues exist, $\sigma=\pm i \omega$, that gives oscillating amplitudes of the form  $exp({\pm i \omega \tau^{(1)}})$. Returning to the original log-time $ \tau$,   it gives  high frequency contributions of the form $ {\bf{U}}^{(1)} = A \exp({ \pm i \lambda \omega \tau}) $. 
We assume that, besides the two eigenmodes in the kernel of ${\mathcal{M}}_{0}$ with $ \omega = 0$, there are at least two other modes of oscillations with two different eigenfrequencies, $\omega_1$ and $\omega_2$, and two amplitudes $A_1$ and $A_2$. The scaling of the oscillations with respect to the big parameter $\lambda$ is 
\begin{equation}
\omega_1 \sim  \omega_2 \sim 0(1)  \textrm{,} \qquad  \mathrm{and}  \qquad A_1\sim A_2\sim 0(1) 
 \textrm{.}
\label{eq:omega}
\end{equation}

The next subsection is about a variational formulation of the eigenvalue problem. Afterwards we come back to the formulation of the solvability condition by taking into account the oscillations and pursuing the expansion of (\ref{eq:Euler3sm}) at order $\lambda$. 

\subsection{Variational formulation of the linearized problem}

Because Euler's equations have a variational formulation (in Lagrangian coordinates) the same must be true for the problem derived by linearization around a non-trivial steady solution of Euler's equations. A derivation of this formulation is given below. 

Let us begin with the simple partial differential equation: 
$$  \frac{\partial a}{\partial t} =   \frac{\partial a}{\partial x} \textrm{,}$$
where $a(x, t)$ is a smooth complex valued function of two real variables, $x$ and $t$. This simple equation is the Euler-Lagrange equation following from the condition of stationarity of the real valued functional: 
$$ {\mathcal{L}} =  \frac{i}{2} \ \int {\textrm{d}}x  \   {\textrm{d}}t \left((a a_t^* - a^* a_t) - (a a_x^* - a^* a_x) \right)  \textrm{,}$$
where $a^*$ is the complex conjugate of $a$ and indices $t$ and $x$ are for derivatives with respect to $t$ and $x$. 
If one assumes that $a$ depends on time like $a = e^{i  \omega t} A(x)$, the function $A(x)$ is derived by the condition of stationarity of the functional: 
$$ {\mathcal{L}}' =  \frac{1}{2}  \int {\textrm{d}}x  \ \left(2\omega A A^* - i (A A_x^* - A^* A_x) \right)  \textrm{.}$$
The principle of this calculation can be extended to get the variational formulation of the problem of eigenvalues and eigenfunctions of the operator ${\mathcal{M}}_{0}$ defined in (\ref{eq:Mo}). 
This operator, denoted  ${\mathcal{M}}$  for simplicity in this subsection, is defined by its action on a vector field ${\bf{V}}$. The result of this action is another vector given by
 \begin{equation} 
 \left({\mathcal{M}}{\bf{V}}\right)_i  = V_j \partial_jU_i^{(0)} + U_j^{(0)} \partial_j V_i + \partial_i  P 
 \textrm{.}
 \label{eq:MV.1}
\end{equation}   
where $U_i^{(0)} $ is a given vector field of zero divergence, $i$ being a discrete coordinate index and where the summation on repeated indices is done. Moreover $\partial_j$ is for the partial derivative with respect to coordinate $r_j$. We are looking for a variational formulation of the eigenvalue problem 
 \begin{equation}
 ({\mathcal{M}}{\bf{V}})_i = i \omega {\bf{V}}_i   \textrm{,}
  \label{eq:MV.3}
\end{equation}   
where ${\bf{V}}$ is divergence less.  That  $ \omega$ is real follows from the property of conservation of the kinetic energy by the linearized problem. As done in the example, we take a complex valued vector field $ {\bf{V}}$, as well as a complex pressure. 
The pressure $P$ in the above equations is a Lagrange multiplier allowing to impose incompressibility of the ${\bf{V}}$ field.

Taking inspiration from the example above, one finds that the eigenvalue equation together with the incompressibility condition can be derived from an Euler-Lagrange condition for the fields ${\bf{V}}$, $P$ and their complex conjugate ${\bf{V}}^*$ and $P^*$. This condition makes stationary the following  real valued functional: 
 \begin{equation}
  {\mathcal{L}} =  \frac{1}{2} \ \int {\textrm{d}}{\bf{r}}  \ \left(- \omega V_i V_i^*+ i (V_i^* U^{(0)}_j ( \partial_j V_i) - V_i U^{(0)}_j ( \partial_j V_i^*))+  \frac{i}{2} (V_i \partial_i P^* -  V_i^* \partial_i P)\right)  
 \textrm{.}
 \label{eq:MV.2}
\end{equation}   

The pressure field $P$ and its complex conjugate can be expressed as a linear transform of the velocity fields $V_i$ and $V_i^*$ by taking the divergence of equation   (\ref{eq:MV.3}) and solving the resulting Poisson equation for $P$ under the condition that $P$ decreases at infinity. Taking the gradient of the resulting expression one finds a contribution to the gradient of $P$ that is an explicit linear transform of $V$. 

Because the incompressibility condition induces a non local term in the equation for the eigenvalue problem, it seems hard to get theoretical results for this eigenvalue problem. One would like to know the structure of this spectrum for a localized velocity field ${\bf{U}}^{(0)}$. By analogy with standard Sturm-Liouville equations  one might guess that this spectrum is discrete, but this is not a very strong argument. Extending the Sturm-Liouville theory to the present case does not look very easy, in particular because the operator ${\mathcal{M}}$ is not self adjoint and non local.  

\subsection{Solvability condition}
We shall now consider the next order terms of the expansion of (\ref{eq:Euler3sm}) in power of $1/\lambda$, which are of order $\lambda$. This part is relevant because  it  justifies the above scaling for the time as $\tau=\tau^{(1)}/\lambda$, and prove that the expansion performed around ${\bf{U}}^{(0)} $ cannot give a {\textit{bona fide}} solution without considering the time derivative term.
 Introducing (\ref{eq:expansion}) in (\ref{eq:Euler3sm}) and gathering  the terms of  order $\lambda$,  we obtain, 
\begin{equation}
{\mathcal{M}}_{0} [ {\bf{U}}^{(2) },P^{(2)} ] =-{\bf{U}}^{(1)} \cdot \nabla {\bf{U}}^{(1)} + \alpha {\bf{U}}^{(0)} + \beta{\bf{R}}\cdot \nabla {\bf{U}}^{(0)} 
 \textrm{,}
\label{eq:ordre2}
\end{equation}
In (\ref{eq:ordre2}) the operator  ${\mathcal{M}}_{0}$ is defined as above by the relation 
\begin{equation}
{\mathcal{M}}_{0} [ {\bf{U}}^{(2) },P^{(2)} ]= {\bf{U}}^{(2)}\cdot \nabla {\bf{U}}^{(0)} + {\bf{U}}^{(0)}\cdot \nabla {{\bf{U}}}^{(2)} +  \nabla P^{(2)} 
 \textrm{,}
\label{eq:Mo2}
\end{equation}
 where  the pressure $ P^{(2)} $ allows to satisfy the incompressibility condition  $\nabla \cdot {\bf{U}}^{(2)} =0$.

The existence or not of solution of  equation (\ref{eq:ordre2}) for ${\bf{U}}^{(2)}$ depends if operator $ {\mathcal{M}}_{0}$
has zero as eigenvalue (namely if it has or not a non empty kernel) and if the right-hand side of equation (\ref{eq:ordre2}) has or not components along the corresponding eigenvector(s). If it has, this equation has no solution.  We show  first that $ {\mathcal{M}}_{0}$ has a non-empty kernel. This follows from two continuous symmetries of (\ref{eq:Euler6ss}), which is our starting point at leading order,  defined above  as the Euler part of the Euler-Leray steady equation. Those two symmetries are first the dilation invariance of the steady Euler equation: if ${\bf{u}}({\bf{r}})$ is a solution, then $\mu {\bf{u}}({\bf{r}})$ is also a solution for an arbitrary real $\mu$, this assuming of course that there is no boundary (an assumption made in this work). From this invariance with $\mu$ close to $1$ one obtains ${\mathcal{M}}_{0} [{\bf{U}}^{(0)},P^{(2)}] = 0$ if $P^{(2)}  = 2P^{(0)}$. 

A second element of the kernel is derived from the spatial dilation invariance of the steady Euler equations. The property that   ${\bf{u}}(\mu {\bf{r}})$ is a solution for an arbitrary real $\mu$  if ${\bf{u}}({\bf{r}})$ is a solution, can be used for $\mu$ close to $1$. One obtains $ {\mathcal{M}}_{0} [{\bf{R}}\cdot \nabla {\bf{U}}^{(0)}] = 0$. This follows by making operate ${\bf{R}}\cdot \nabla $ on the Euler equation for ${\bf{U}}^{(0)}$.  In summary we find that $ {\mathcal{M}}_{0}$ has at least two eigenvectors in his kernel, which are $\left({\bf{U}}^{(0)},P^{(0)}/2\right)$ and $\left({\bf{R}}\cdot \nabla {\bf{U}}^{(0)}, {\bf{R}}\cdot \nabla P^{(0)}/2\right)$. 

There are other continuous symmetries of the Euler equation, the geometrical symmetries under rotation and translation and the Galilean invariance. The symmetry under rotation are also shared by the advection term and so are irrelevant, whereas the Galilean invariance does not bring a solvability condition because of the assumed mirror symmetry of the solution of the Bragg-Hawthorne equations with respect to the plane $z = 0$. 

Therefore we are left with a kernel of ${\mathcal{M}}_{0}$ with two independent vectors, so that equation (\ref{eq:ordre2}) has a solution only if the right-hand side has no component in the kernel of the operator adjoint of $ {\mathcal{M}}_{0}$,  ${\mathcal{M}}_{0}^{\dagger}$  which is a tensor. 

At this stage, before searching an expression for the solvability condition let us  consider what happens for  a  time independent problem, namely  for a spatial perturbation of the  BH solution in the case of Euler-Leray equations (not the modified  equation (\ref{eq:Euler2ss}) which includes the time derivative term).   In this case,  the first order term ${\bf{U}}^{(1)}$ in the expansion (\ref{eq:expansion})  is absent and  at  the new ''first order'' , keeping the same notations as above, we obtain   $${\mathcal{M}}_{0} [ {\bf{U}}^{(2) },P^{(2)} ]= \alpha {\bf{U}}^{(0)} + \beta{\bf{R}}\cdot \nabla {\bf{U}}^{(0)}  $$ in place of (\ref{eq:ordre2}).   A solution for $ ({\bf{U}}^{(2) },P^{(2)})$ exists only if  the r.h.s. of this equation is orthogonal to the kernel of the adjoint operator ${\mathcal{M}}_{0} ^{\dagger}$. But such a condition cannot be realized in a problem without any free parameter, as it is the case if the term $-{\bf{U}}^{(1)} \cdot \nabla {\bf{U}}^{(1)}$ has disappeared in (\ref{eq:ordre2}).  Whence, without the time derivative  term in (\ref{eq:Euler2ss}), the perturbation expansion fails already at first order  to give a {\emph{bona fide}} solution. This is fully consistent with the result of section  \ref{solaxis} showing that, in this axisymmetric geometry, there is no solution of the  {\emph{steady}} Euler-Leray equations, a result already presented in reference  \cite{YP}. In the same reference it was attempted to replace the steady Euler-Leray equation by its time-dependent version (\ref{eq:Euler3sm}) and to build a solution oscillating periodically with respect to the time $\tau$. However this solution was non explicit and it was hard to devise a numerical schema showing its existence. In summary we show that the present problem  the oscillation is necessary to get a free parameter allowing to satisfy the solvability condition that we shall develop now.

We shall take the simplest possible inner product between vector fields depending on ${\bf{R}}$, namely the volume integral of the function defined by the scalar product of two vectors fields. This product between two vector fields, denoted as $<{\bf{U}} \cdot {\bf{W}}> $, 
allows to define the adjoint operator by doing partial integrations. This inner product is defined  for a pair of vector fields ${\bf{V}}$ and ${\bf{W}}$
decaying each sufficiently fast at large distances to make converge the volume integral of their scalar product. Let $V_i$ and $W_j$ be the Cartesian components of the two vector fields (to simplify the discussion we use Cartesian coordinates instead of the original cylindrical ones). 
Dropping the subscript $0$ for the operators ${\mathcal{M}}_{0}$ and  ${\mathcal{M}}_{0}^{\dagger}$, the adjoint operator is defined by the relation,
 \begin{equation} 
  <\left({\mathcal{M}}{\bf{V}} \right)_i  W_i  > = < V_i \left({\mathcal{M}}^{\dagger}{\bf{W}} \right)_i > \textrm{,}
   \label{eq:Mdag}
\end{equation}   
where the summation on like indices is implied for any suitable pair $(V, W)$ making the volume integral converging.  From the definition of ${\mathcal{M}}$ as given in equation (\ref{eq:Mo2}),
the action of ${\mathcal{M}}$ is given by,
 \begin{equation} 
 \left({\mathcal{M}}{\bf{V}}\right)_i  = V_j \partial_jU_i^{(0)} + U_j^{(0)} \partial_j V_i + \partial_i  P^{(2)} 
 \textrm{.}
 \label{eq:MV}
\end{equation}   
Assuming that ${\bf{W}}$ is divergenceless, the pressure term is eliminated in the scalar product, then we get,
 \begin{equation} 
  <\left({\mathcal{M}}{\bf{V}} \right)_i  W_i  > = \int{W_{i} V_j   \partial_jU_i^{(0)} }   +   \int{ W_{i} U_j^{(0)} \partial_j V_i}  
 \textrm{.}
 \label{eq:MiVi}
\end{equation}   
  
The matrix elements of  the adjoint are obtained by interchanging the index $i,j$ in the first term, and  integrating by parts the second term which  becomes $-   \int{ V_{i} U_j^{(0)} \partial_j W_i}   $, it gives
 \begin{equation} 
  \left({\mathcal{M}}^{\dagger}{\bf{W}} \right)_i   =W_{j}   \partial_i U_j^{(0)}    -  U_j^{(0)} \partial_j W_i
      \textrm{.}
       \label{eq:Madjoint}
\end{equation}

Let us now  give an expression of the  solvability condition necessary for  solving (\ref{eq:ordre2}), within the frame (\ref{eq:omega}) where the first order amplitude $ {\bf{U}}^{(1)}$ is,
  \begin{equation} 
 {\bf{U}}^{(1)}= A_{1}{\bf{f}}_{1}({\bf{R}}) e^{(i  \lambda \omega_{1} \tau)}+  A_{2}{\bf{f}}_{2}({\bf{R}}) e^{(i  \lambda \omega_{2} \tau)}+ c.c
  \textrm{.}
  \label{eq:2modes}
\end{equation}

 In the  term ${\bf{U}}^{(1)} \cdot \nabla {\bf{U}}^{(1)}$  of (\ref{eq:ordre2})
 we have first terms oscillating at frequencies $\lambda$ times  $2\omega_1$, $2 \omega_2$ and $(\omega_1 \pm \omega_2)$, and
also nonlinear terms proportional to  the square of the amplitude  $\vert A_1\vert^2$ and $\vert A_2\vert^2$.
 The latter terms have zero frequency and are to be added to the Leray advection term computed with the  velocity field ${\bf{U}}^{(0)}$ to contribute to the solvability condition. We recall that this advection term, which linear with respect to
 ${\bf{U}}^{(0)}$, appears at order $\lambda$ of the expansion  (\ref{eq:expansion}),  and $A_1$ and  $A_2$  appears at order $\lambda^{1/2}$.  The time averaging leading to overlined quantities makes disappear the oscillating terms which are all high frequency. We obtain 
\begin{equation} 
   \overline{{\bf{U}}^{(1)}\cdot \nabla {{\bf{U}}^{(1)}}} = \vert A_1\vert^2  {\bf{F}}_1({\bf{R}})+ \vert A_2\vert^2 {\bf{F}}_2 ({\bf{R}}) \textrm{,}
 \label{eq:moysurt}
\end{equation}
   where the velocity fields  ${\bf{F}}_1({\bf{R}})$ and ${\bf{F}}_2({\bf{R}})$
are computed by averaging  ${\bf{U}}^{(1)}\cdot \nabla {{\bf{U}}^{(1)}}$ over the oscillations at frequencies $\omega_1$ and $\omega_2$. We did not mention another contribution to ${\bf{F}}_1$ coming from the pressure term. This one includes also a contribution quadratic with respect to  $A_1$ and  $A_2$, derived by taking the divergence of the Euler-Leray first equation and by solving the resulting Poisson equation. This does not change fundamentally the calculation and can be included without too much trouble into it.

 To simplify the algebra we have to associate indices to each one of the two zero eigenmodes of ${\mathcal{M}} $. We shall take $a$ to denote the eigenmode ${\bf{U}}^{(0)}$ and  $b$ to the eigenmode ${\bf{R}}\cdot \nabla{\bf{U}}^{(0)}$. Let us denote the two adjoint elements of the kernel as ${\bf{V}}_a$ and  ${\bf{V}}_b$. Therefore the two solvability conditions become, 
 \begin{equation}
\vert A_1\vert^2  < {\bf{V}}_a \cdot {\bf{F}}_1({\bf{R}})> + \vert A_2\vert^2 < {\bf{V}}_a \cdot {\bf{F}}_2({\bf{R}})> = <  {\bf{V}}_a \cdot (\alpha \ {\bf{U}}^{(0)} + \beta\ {\bf{R}}\cdot \nabla {\bf{U}}^{(0)}) > = 0
 \textrm{,}
\label{eq:EulerSolva}
\end{equation}
 \begin{equation}
\vert A_1\vert^2  < {\bf{V}}_b \cdot {\bf{F}}_1({\bf{R}})> + \vert A_2\vert^2 < {\bf{V}}_b \cdot {\bf{F}}_2({\bf{R}})> =  <  {\bf{V}}_b \cdot (\alpha\ {\bf{U}}^{(0)} + \beta\ {\bf{R}}\cdot \nabla {\bf{U}}^{(0)}) > = 0
 \textrm{,}
\label{eq:EulerSolvb}
\end{equation}
This makes a set of two linear equations for $\vert A_1\vert^2 $ and $\vert A_2\vert^2 $. Of course one has to worry about the sign of those two quantities. This is something hard to discuss without knowing precisely the values of the coefficients that depend themselves upon the choice of a solution of the Bragg-Hawthorne equation. Besides a numerical approach it seems to be hard to conclude at the existence or not of such a solution with positive values of  $\vert A_1\vert^2 $ and $\vert A_2\vert^2 $. 

Let us notice that by continuing the expansion to higher order with respect to the small parameter $1/\lambda$ one shall meet at every order two solvability conditions. To satisfy those conditions one needs to adjust higher order terms in the expansion of  the amplitude $\vert A_1\vert^2 $ and $\vert A_2\vert^2 $. This gives at least formally the right number of free parameters at every order.

 \section{Conclusion and perspectives}
 \label{Concperspect}
 The aim of this note was to present a possible scenario for proving (in the physical sense) the existence of non trivial solutions of the Euler-Leray equations. It shows in particular that such solution should have a non trivial structure at least in the limit under consideration.  We believe however that some features of this solution will survive in other types of self similar solutions of the Euler-Leray equations. Specifically the oscillations in the log-time can be seen as giving a kind of Stokes drift \cite{stokes} balancing the divergence imposed by the change of coordinates of the Leray transform.

{}
\end{document}